\documentclass[hyper,11pt,paper]{article}
\usepackage{jheppub}
\usepackage[usenames,dvipsnames,table]{xcolor}
\usepackage{graphicx,amsmath,amssymb,multirow,array,bm,mathrsfs}
\usepackage{epsf,amsfonts}
\usepackage[numbers,sort&compress]{natbib}
\usepackage[titletoc,toc,title]{appendix}

\title{A breakdown of a universal hydrodynamic relation in Gauss-Bonnet gravity}
\author{Evgeny Shaverin}

\affiliation{Department of Physics, Technion, Haifa 32000, Israel}
\emailAdd{evgeny@tx.technion.ac.il}

\abstract{We compute the second order transport coefficients of a hydrodynamic theory with Einstein-Gauss-Bonnet gravity dual. We show the breakdown of the universal hydrodynamic relation $ -2 \lambda_0 + 4 \lambda_1 - \lambda_2 = 0 $ for the general values of the Gauss-Bonnet coupling.}

\begin{document}
\maketitle

\section{Introduction} \label{S:Intro}

\subsection*{Conformal viscous hydrodynamics}
Hydrodynamics is a low energy effective description of many-body systems which are studied at length scales much larger than their mean free path. When only energy and momentum are conserved, this effective description is given by a handful of hydrodynamics fields: the velocity field $ u^{\mu} $ (normalized so that $ u^{\mu}u_{\mu}=-1 $), and the temperature field $ T $. The dynamics of these fields is governed by energy conservation\,,
\begin{equation}
	\partial_{\mu}T^{\mu\nu} = 0 \, .
\end{equation}

In a conformal theory, the dependence of the stress-energy tensor $ T^{\mu\nu} $ on the hydrodynamic fields takes the following form
\begin{equation} \label{E:GeneralTmn}
	T^{\mu\nu} = P \left(4 u^{\mu}u^{\nu} + \eta^{\mu\nu}\right) + \tilde{\Pi}^{\mu\nu} \, ,
\end{equation}
where $ P=P(T) $ is the pressure and $ \tilde{\Pi}^{\mu\nu} $  denotes dissipative terms. The tensor $ \tilde{\Pi}^{\mu\nu} $ depends on gradients of the hydrodynamic variables $ T $ and $ u^{\mu} $. It is convenient to impose on $ \tilde{\Pi}^{\mu\nu} $ the Landau frame condition $ u_\mu \tilde{\Pi}^{\mu\nu} = 0 $. Following \cite{Baier:2007ix}, an expansion of $ \tilde{\Pi}^{\mu\nu} $ in derivatives of $ T $ and $ u^{\mu} $ up to second order is given by
\begin{equation} \label{DissipativeParts}
	\tilde{\Pi}^{\mu\nu} = - \eta \sigma^{\mu\nu} + \sum_{i=0}^{4} \lambda_{i} \Sigma^{(i)\mu\nu} \, .
\end{equation}
The coefficient $ \eta $ in \eqref{DissipativeParts} is the {shear} viscosity, and the coefficients $ \lambda_i $, $ i=0,\ldots,4 $, are referred to as `second order transport coefficients'. The tensors $ \sigma^{\mu\nu} $ and $ \Sigma^{(i)}_{\mu\nu} $ are given by:
\begin{align} \label{2orderTemrs}
\begin{split}
	\sigma^{\mu\nu} &= 2\partial^{\langle\mu}u^{\nu\rangle} \, ,\\
	\Sigma^{(0)}_{\mu\nu} &= {}_{\langle}u^{\alpha}\partial_{\alpha}\sigma_{\mu\nu\rangle} + \frac{1}{3} \sigma_{\mu\nu}\partial_{\alpha}u^{\alpha} \, ,\\
	\Sigma^{(1)}_{\mu\nu} &= \sigma_{\langle \mu \alpha}\sigma^{\alpha}{}_{\nu\rangle} \,,
	\quad
	\Sigma^{(2)}_{\mu\nu} = \sigma_{\langle \mu\alpha} \omega^{\alpha}{}_{\nu\rangle} \, , \\
	\quad
	\Sigma^{(3)}_{\mu\nu} &= \omega_{\langle \mu\alpha} \omega^{\alpha}{}_{\nu\rangle} \, ,
	\quad
	\Sigma^{(4)}_{\mu\nu}  = R_{\langle\mu\nu\rangle} - 2 u^{\alpha} R_{\alpha\langle\mu\nu\rangle\beta} u^{\beta} \,,
\end{split}
\end{align}
with the following definitions: $R_{\mu\nu} $ and $ R_{\alpha\mu\nu\beta} $ are the Ricci and Riemann tensor respectively, the vorticity, $ \omega_{\mu\nu} $, is given by $ \omega_{\mu\nu}=\frac{1}{2}P_{\mu}{}^{\alpha}P_{\nu}{}^{\beta}\left(\partial_{\alpha}u_{\beta}-\partial_{\beta}u_{\alpha}\right) $, and triangular brackets denote a traceless and transverse projection
\begin{equation} \nonumber
	A^{\langle \mu \nu \rangle}= \frac{1}{2} P^{\mu\alpha}P^{\nu\beta} \left(A_{\alpha\beta}+A_{\beta\alpha}\right) - \frac{1}{3}P^{\mu\nu}P^{\alpha\beta}A_{\alpha\beta} \, ,
\end{equation}
where we have defined the projection on the spatial directions $ P^{\mu\nu} = \eta^{\mu\nu}+u^{\mu}u^{\nu} $. We assume that the field theory is living in flat space, where the Ricci and Riemann tensors vanish, therefore $ \Sigma^{(4)}_{\mu\nu} = 0 $. Collecting (\ref{E:GeneralTmn}) and (\ref{DissipativeParts}) we find that in our setup,
\begin{equation} \label{E:GeneralTmn1}
	T^{\mu\nu} = P \left(4 u^{\mu}u^{\nu} + \eta^{\mu\nu}\right) - \eta \sigma^{\mu\nu} + \sum_{i=0}^{3} \lambda_{i} \Sigma^{(i)\mu\nu} \, .
\end{equation}

In \cite{Kovtun:2003wp} it was conjectured that for all known physical systems the ratio of the shear viscosity to entropy density is given by
\begin{equation} \label{BOUND}
	\frac{\eta}{s} \ge \frac{1}{4 \pi} \, .
\end{equation}
The bound \eqref{BOUND} is found to be saturated in all thermally equilibrated systems with isotropic and homogeneous configuration described by a dual two-derivative gravity action \cite{Buchel:2003tz,Kovtun:2004de,Buchel:2004qq}. The bound \eqref{BOUND} has been shown to be violated in the non-isotropic configurations, see \cite{Erdmenger:2010xm,Basu:2011tt,Erdmenger:2011tj,Rebhan:2011vd,Jain:2014vka,Ovdat:2014ipa}. Recent investigations \cite{Kats:2007mq,Brigante:2007nu,Buchel:2008vz} show that in theories dual to higher derivative gravity there is a violation of this bound. In the special case of Gauss-Bonnet gravity,
\begin{equation} \label{S_EGB}
	\mathcal{S}_{\rm{EGB}} = -\frac{1}{16 \pi G_5} \int \sqrt{-g} \left( R + \frac{12}{L^2} - \theta L_{GB} \right) d^5x \, ,
\end{equation}
\begin{equation}\nonumber
	L_{GB} = R_{mnpq}R^{mnpq} - 4 R_{mn}R^{mn} + R^2 \, ,
\end{equation}
it was shown in \cite{Brigante:2007nu} that for general values of a Gauss-Bonnet coupling $ \theta $
\begin{equation} \label{eta/s}
	\frac{\eta}{s} = \frac{1}{4 \pi} \left( 1-8\theta \right) \, .
\end{equation}

When considering the second order transport coefficients, it has been observed that the linear relation
\begin{equation} \label{relation}
	-2 \lambda_0 + 4 \lambda_1 - \lambda_2 = 0 \,,
\end{equation}
seems to be satisfied for a large class of two derivative gravity theories  \cite{Erdmenger:2008rm}. In contrast, the coefficients $\lambda_i$ where also computed perturbatively using  kinetic theory where the relation (\ref{relation}) seems not to hold.

Surprisingly, Gauss-Bonnet gravity seems not to correct the relation \eqref{relation} when it is treated perturbatively \cite{Shaverin:2012kv}. This result has been confirmed and extended in \cite{Grozdanov:2014kva} to include perturbatively small eight derivative corrections. In what follows we extend these results and compute the expression \eqref{relation} to all orders in the Gauss-Bonnet coupling. We find that 
\begin{equation} \label{FinResult}
	- 2 \lambda_0 + 4 \lambda_1  - \lambda_2 = \frac{\pi^{2}T^{2}}{\pi G_{5}}\left[\frac{\sqrt{2}\left(1-\sqrt{1-8\theta}+16\theta\right)\theta}{\left(1+\sqrt{1-8\theta}\right)^{3/2}}\right] \, .
\end{equation}
Thus, the relation \eqref{relation} which seems to hold even at the perturbative level in 2, 4 and 8 derivative gravity breaks down at the non perturbative level. We note that the relation \eqref{FinResult} has been published before in \cite{Grozdanov:2014kva} using methods and the non-perturbative expressions for all Gauss-Bonnet transport coefficients published in \cite{Andrei}.
Treating the Gauss-Bonnet coupling $ \theta $ non perturbatively via AdS/CFT may be regarded as a toy model for probing finite $ N $ corrections in a gauge theory \cite{Brigante:2007nu,Kats:2007mq,Buchel:2008vz}.

Our work is organised as follows. In the chapter 2 we introduce the action and equations of motion for the Einstein-Gauss-Bonnet gravity, dual to a viscous fluid in the field theory. Using black hole solution to the equations of motion and AdS/CFT dictionary we find an ideal fluid (non-dissipative) part of the energy momentum tensor. In the chapter 3 we introduce derivative corrections to the gravity dual corresponding to the viscous contributions in the field theory side. Solving the equations of motion up to second order in derivatives we reproduce the known result for the shear viscosity over entropy density (\ref{eta/s}) and show explicitly the breakdown of the universal relation (\ref{FinResult}).

\section{AdS Gauss-Bonnet black holes}
The equations of motion for Gauss-Bonnet gravity can be derived from the following action,
\begin{equation} \label{action}
	\mathcal{S} = \mathcal{S}_{\rm{EGB}} + S_{\rm{b}}
\end{equation}
where $ \mathcal{S}_{\rm{EGB}} $ was defined in (\ref{S_EGB}) and $S_{\rm{b}}$ are appropriate counter terms, imposed to make the variational principle well defined \cite{PhysRevD.15.2752, Balasubramanian:1999re}. The Roman indices $m,\,n=0,\ldots,4$ refer to bulk (AdS) quantities while Greek indices $\mu,\,\nu = 0,\ldots,3$ refer to boundary quantities. We set $ L=1 $ from now on. The equations of motion that follow from (\ref{action}) are given by
\begin{multline}
\label{EOM}
	R_{mn}-\frac{1}{2}Rg_{mn}-6g_{mn}-\frac{\theta}{2}g_{mn}L_{GB} \\
	+2\theta\left(R_{mpql}R_{n}^{\phantom{n}pql}-2R^{pq}R_{mpnq}-2R_{m}^{\phantom{m}q}R_{qn}+RR_{mn}\right)=0 \,.
\end{multline}

The black hole solution to \eqref{EOM} is given by 
\begin{equation} \label{eq:ansatz0}
	ds^2 = -r^2 f(\rho) dt^2 + r^2 \left(dx^i\right)^2 + 2 s(\rho) dt dr \, ,
\end{equation}
where $ \rho\equiv br $ and
\begin{align}\label{0th_order_functions}
\begin{split}
	s(\rho) & = \sqrt{\frac{1+\sqrt{1-8\theta}}{2}}\equiv S_{0}\\
	f(\rho) & = \frac{S_{0}^{2}}{4\theta}\left(1-\sqrt{1-8\theta\left(1-\frac{1}{\rho^4}\right)}\right) \, ,
\end{split}
\end{align}
with $ b $ an integration constant associated with the mass of the black hole \cite{PhysRevLett.55.2656,Cai:2001dz}.

Under the gauge-gravity duality, the black hole solution is dual to a thermal state of the black hole. The energy momentum tensor of this state can be computed using the standard prescription, \cite{Brihaye:2008kh,Dutta:2008gf,Buchel:2008vz},
\begin{flalign}
\label{E:Tmnprescription}
	& T_{\mu\nu}   = 
		\lim_{r\to\infty}\frac{r^{2}}{8\pi G_5}
		\left(\mathcal{K}_{\mu\nu}-\mathcal{K}\gamma_{\mu\nu}
		+2\theta\left(3\mathcal{J}_{\mu\nu}-\mathcal{J}\gamma_{\mu\nu}\right)
		-3\gamma_{\mu\nu}+\left[\frac{-2S_{0}^{2}+3S_{0}-1}{S_{0}}\right]\gamma_{\mu\nu}\right)
\end{flalign}
where:
\begin{itemize}
	\item $ \gamma_{mn} = g_{mn} - N_m N_n $ is the boundary metric

	\item $ N_n = \delta_n^r / \sqrt{g^{rr}} $ a unit outward vector to the boundary

	\item $ \mathcal{K}_{mn} = -\frac{1}{2} \left(\nabla_{m} N_n + \nabla_n N_m\right) $ is the extrinsic curvature on the boundary (the covariant derivative here is with respect to the bulk metric $ g_{mn} $)

	\item $ \mathcal{J}_{mn} = \frac{1}{3} \Big(2 \mathcal{K} \mathcal{K}_{mp}\mathcal{K}^p_{\phantom{p}n} + \mathcal{K}_{ps}\mathcal{K}^{ps} \mathcal{K}_{mn} - 2 \mathcal{K}_{mp}\mathcal{K}^{ps}\mathcal{K}_{sn} - \mathcal{K}^2 \mathcal{K}_{mn}\Big) $

	\item $\mathcal{K}$ and $\mathcal{J}$ are the trace (using $ \gamma_{mn} $ to contract indices) of $\mathcal{K}_{mn}$ and $\mathcal{J}_{mn}$ respectively

	\item $ S_{0} =  \sqrt{\frac{1+\sqrt{1-8\theta}}{2}}$
\end{itemize}
Inserting \eqref{eq:ansatz0} into \eqref{E:Tmnprescription} and using the following relation
\begin{equation}
	T = \frac{S_0}{b \pi}
\end{equation}
between the temperature of the dual theory and the horizon parameter $b$, derived in (\ref{TEMP}), we find
\begin{equation} \label{zeroth_order}
	T^{\mu\nu} = \frac{T^4 \pi^4}{16\pi G_{5} S_0^3} \left(4 u^{\mu}u^{\nu} + \eta^{\mu\nu}\right) \, .
\end{equation}
The parameter $G_5$ is theory dependent. For instance, in the planar limit of strongly coupled $\mathcal{N}=4$ super Yang Mills one finds
\begin{equation}
	G_5 = \frac{\pi}{2N^2}\,.
\end{equation}
Gauss-Bonnet coupling $ \theta $ can be regarded as $ \mathcal{O}(N) $ corrections to the action \cite{Kats:2007mq,Buchel:2008vz}.

\section{Solving the equations of motion perturbatively}
In order to compute the transport coefficients of the fluid dual to the theory defined by \eqref{action} we use the method of \cite{Bhattacharyya:2008jc}; The metric (\ref{eq:ansatz0}) may be boosted with velocity $ u^{\alpha} = \frac{1}{\sqrt{1-\beta^2}} \left( 1, \vec{\beta} \right) $, resulting in the line element
\begin{equation} \label{eq:ansatz1}
	ds^2 = -r^2 f(br) u_{\mu}u_{\nu}dx^{\mu}dx^{\nu} + r^2 P_{\mu\nu} dx^{\mu}dx^{\nu} - 2 s(br) u_{\mu} dx^{\mu} dr \, .
\end{equation}
We now promote $ u^\mu $ and $ b $ to be a slowly-varying fields,
\begin{equation} \label{coordep}
	\beta_i \rightarrow \beta_i(x^{\alpha}) \, , \quad b \rightarrow b(x^{\alpha}) \,.
\end{equation}
Inserting (\ref{coordep}) into (\ref{eq:ansatz1}) is in general, not a solution to the equations of motion (\ref{EOM}). Therefore we correct (\ref{eq:ansatz1}) by adding new terms to the metric. The corrections to the metric may be classified as follows. Let us decompose the metric into scalar, vector and tensor modes of the $SO(3)\subset SO(3,1)$ symmetry under which $u^{\mu}$ is (locally) invariant, i.e.,
\begin{flalign}
\label{E:lineelement}
	ds^{2} & = r^{2} k u_{\mu}u_{\nu}dx^{\mu}dx^{\nu}
		+r^{2}P_{\mu\nu}dx^{\mu}dx^{\nu}-2Su_{\mu}dx^{\mu}dr \nonumber \\
		& \phantom{=}+r^{2}\left(u_{\mu}V_{\nu}+u_{\nu}V_{\mu}\right)dx^{\mu}dx^{\nu}
		+r^{2}\Pi_{\mu\nu}dx^{\mu}dx^{\nu}
\end{flalign}
and expand $k$, $S$, $V_{\mu}$ and $\Pi_{\mu\nu}$ in gradients of $ u^{\alpha} $ and $ b $. It is convenient to denote the $n$'th order correction to $k$, $S$, $V_{\mu}$ and $\Pi_{\mu\nu}$ with a superscript $(n)$. For instance,
\begin{flalign} \label{kSVP}
\begin{split}
	k & = k^{(0)} + k^{(1)} + k^{(2)} + \ldots = - f(\rho) + k^{(1)} + k^{(2)} + \ldots  \\
	S & = S^{(0)} + S^{(1)} + S^{(2)} + \ldots = s(\rho) + S^{(1)} + S^{(2)} + \ldots  \\
	V_{\mu} & = V^{(0)}_{\mu} + V^{(1)}_{\mu} + V^{(2)}_{\mu} + \ldots = V^{(1)}_{\mu} + V^{(2)}_{\mu} + \ldots  \\
	\Pi_{\mu\nu} & = \Pi^{(0)}_{\mu\nu} + \Pi^{(1)}_{\mu} + \Pi^{(2)}_{\mu} + \ldots = \Pi^{(1)}_{\mu} + \Pi^{(2)}_{\mu} + \ldots ,
\end{split}
\end{flalign}
where $ f(\rho) $ and $ s(\rho) $ are given in (\ref{0th_order_functions}). Inserting (\ref{kSVP}) into (\ref{E:lineelement}) and using the equations of motion (\ref{EOM}) we find, at each order in $ n>0 $,
\begin{flalign}
\begin{split}
\label{E:EOMs}
	\left(S^{(n)}\right)^{\prime} & = \mathbf{S}^{(n)}\\
	\left( \rho^4 \left(S_0^2-4\theta f(\rho)\right) \left(k^{(n)}+ \frac{2 f(\rho)}{S_0} S^{(n)} \right)  \right)^{\prime} 
		& = \mathbf{k}^{(n)} \\
	\left( \rho^5 \left(S_0^2-4\theta f(\rho)\right)\, V_{\mu}^{(n)\,\prime} \right)^{\prime} 
		& = \mathbf{V}_{\mu}^{(n)} \\
	\left( \frac{\rho^5 f(\rho)}{S_0^2-4\theta f(\rho)} \,
		\Pi_{\mu\nu}^{(n)\,\prime}\right)^{\prime} & =\mathbf{P}_{\mu\nu}^{(n)} \, ,
\end{split}
\end{flalign}
where the quantities on the right-hand side of (\ref{E:EOMs}) depend on the known $ 0 $ to $ n-1 $ order metric components. Integrating (\ref{E:EOMs}) leads to
\begin{flalign}
\begin{split}
\label{E:SOL}
	S^{(n)} & = - \int_{\rho}^{\infty} \mathbf{S}^{(n)}(x')dx'\\
	k^{(n)} & = 
		\frac{\int_{1}^{\rho} \mathbf{k}^{(n)}(x')dx'}{\rho^4 \left(S_0^2-4\theta f(\rho)\right)} 
		- \frac{2 f(\rho)}{S_0} S^{(n)} + \frac{\mathcal{C}^1}{\rho^4 \left(S_0^2-4\theta f(\rho)\right)}\\
	V_{\mu}^{(n)} & = - \int_{\rho}^{\infty}
		\frac{\int_{1}^x \mathbf{V}_{\mu}^{(n)}(x')dx'}{x^5 \left(S_0^2-4\theta f(x)\right)}dx
			+\mathcal{C}_{\mu}^2 \int_{\rho}^{\infty}\frac{dx}{x^5 \left(S_0^2-4\theta f(x)\right)}\\
	\Pi_{\mu\nu}^{(n)} & = - \int_{\rho}^{\infty} \frac{\left(S_0^2-4\theta f(x)\right) \int_{1}^x \mathbf{P}_{\mu\nu}^{(n)}(x')dx'}{x^5 f(x)}dx \, .
\end{split}
\end{flalign}
The boundary conditions for the integral were chosen to ensure that
\begin{enumerate}
\item The boundary metric is flat: $ \lim\limits_{r\rightarrow\infty} ds^2 = r^2 \eta_{\mu\nu} dx^{\mu}dx^{\nu} $, i.e. for $ n\ge1 $:
\begin{flalign} \nonumber
	\lim_{r\rightarrow\infty}k^{(n)} & = 0 \, , \quad
	\lim_{r\rightarrow\infty}V_{\mu}^{(n)} = 0 \, , \quad
	\lim_{r\rightarrow\infty}\Pi_{\mu\nu}^{(n)}  = 0 \, .
\end{flalign}
\item The metric is regular at $ r=1 $.
\item $ T^{\mu\nu} $ is in the Landau frame: there are no $ 1/\rho^4 $ dependence in the near-boundary expansion of $ k^{(n)} $ and $ V_{\mu}^{(n)} $ with $ n\ge1 $.
\end{enumerate}

With a perturbative solution to the equations of motion at hand one can use the prescription \eqref{E:Tmnprescription} to compute the energy momentum tensor order by order in the derivative expansion.

\subsection{First order}
Using (\ref{kSVP}) to first order in derivatives, we find (\ref{E:EOMs}) with
\begin{align}
\begin{split}
\label{E:sources1}
	\mathbf{S}^{(1)} & = 0\\
	\mathbf{k}^{(1)} & = b \, \left(\frac{2}{3} S_0 \rho^{3 }\left(S_0^2-4\theta f(\rho)\right)\right)^{\prime} \partial_{\alpha}u^{\alpha}\\
	\mathbf{V}_{\mu}^{(1)} & = b \left(S_0 \rho^3 \left(S_0^2-4\theta f(\rho)\right)\right)^{\prime} u^{\alpha}\partial_{\alpha}u_{\mu}\\
	\mathbf{P}_{\mu\nu}^{(1)} & = b \left(\frac{- S_0 \rho^3}{S_0^2-4\theta f(\rho)}\right)^{\prime} \sigma_{\mu\nu} \, .
\end{split}
\end{align}
Inserting (\ref{E:sources1}) into (\ref{E:SOL}) we get
\begin{flalign}
\begin{split}
\label{1st_order_functions}
	S^{(1)} & = 0 \\
	k^{(1)} & = b\frac{2 S_0}{3 \rho} \, \partial_{\alpha}u^{\alpha}\\ 
	V_{\mu}^{(1)} & = - b \frac{S_0}{\rho}u^{\alpha} \, \partial_{\alpha}u_{\mu} \\
	\Pi_{\mu\nu}^{(1)} & = b \, \pi(\rho) \, \sigma_{\mu\nu} \, ,
\end{split}
\end{flalign}
where
\begin{equation} \nonumber
	\pi(\rho) = - \int_{\rho}^{\infty} \frac{\left(S_0^2-4\theta f(x)\right)-S_0^2 x^3}{S_0 x^5 f(x)} \, dx \, .
\end{equation}

Evaluating (\ref{E:Tmnprescription}) by using the first order metric solution and temperature relation (\ref{TEMP}), we get
\begin{equation} \label{eta}
	T^{\mu\nu}=\frac{T^{4}\pi^{4}}{16\pi G_{5}S_{0}^{3}}\left(4u^{\mu}u^{\nu}+\eta^{\mu\nu}\right)-\frac{T^{3}\pi^{3}\left(1-8\theta\right)}{16\pi G_{5}S_{0}^{3}}\sigma^{\mu\nu} \, .
\end{equation}
Using \eqref{E:GeneralTmn1} we can read the shear viscosity:
\begin{equation}
	\eta = \frac{T^{3}\pi^{3}\left(1-8\theta\right)}{16\pi G_{5}S_{0}^{3}} \, .
\end{equation}
Note that using $ s = \frac{dP}{dT} $ together with the relations (\ref{0th_order_functions}), we recover the result of \cite{Brigante:2007nu}
\begin{equation}
	\frac{\eta}{s} = \frac{1}{4\pi} \left(1-8\theta\right) \, .
\end{equation}

\subsection{Second order}
Inserting the zeroth and first order solutions (\ref{0th_order_functions}) and (\ref{1st_order_functions}) into (\ref{kSVP}) and solving (\ref{EOM}), one finds that the solution to the equations of motion take the form
\begin{flalign} \label{secondordermetricC}
\begin{split}
	S & = S_0 + \tilde{\mbox{s}}_{4}(\rho)\,\mathfrak{S}_{4}+\tilde{\mbox{s}}_{5}(\rho)\,\mathfrak{S}_{5}\\
	k & = -f(b r) + b\frac{2 S_0}{3 \rho} \, \partial_{\alpha}u^{\alpha} 
		+ \tilde{\mbox{k}}_{0}(\rho)\,\mathbf{s}_{3} + \tilde{\mbox{k}}_{1}(\rho)\,\mathfrak{S}_{1} \\
		& \quad\, +\tilde{\mbox{k}}_{3}(\rho)\,\mathfrak{S}_{3}+\tilde{\mbox{k}}_{4}(\rho)\,\mathfrak{S}_{4}
		+\tilde{k}_{5}(\rho)\,\mathfrak{S}_{5}\\
	V_{\mu} & = - b \frac{S_0}{\rho}u^{\alpha} \, \partial_{\alpha}u_{\mu} + \tilde{\mbox{w}}_{4}(\rho)\,\mathbf{v}_{4\,\mu}
		+\tilde{\mbox{w}}_{5}(\rho)\,\mathbf{v}_{5\,\mu} \\
		& \quad\, +\tilde{\mbox{v}}_{1}(\rho)\,\mathfrak{V}_{1\,\mu}
			+ \tilde{\mbox{v}}_{2}(\rho)\,\mathfrak{V}_{2\,\mu}+\tilde{\mbox{v}}_{3}(\rho)\,\mathfrak{V}_{3\,\mu}\\
	\Pi_{\mu\nu} & = b \, \pi(\rho) \, \sigma_{\mu\nu} 
		+ b^{2}\sum_{i=0}^{3} \tilde{\pi}_{i}(\rho) \, \Sigma_{\mu\nu}^{\left(i\right)} \, ,
\end{split}
\end{flalign}
where 
\begin{align*}
		\mathbf{s}_{3} & = \frac{1}{b}P^{\alpha\beta}\partial_{\alpha}\partial_{\beta}b,\\
		\mathfrak{S}_{1} & =\mathcal{D}u^{\alpha}\mathcal{D}u_{\alpha},\quad
		\mathfrak{S}_{3}=\left(\partial_{\mu}u^{\mu}\right)^{2}\\
		\mathfrak{S}_{4} & = l_{\mu}l^{\mu},\quad
		\mathfrak{S}_{5}=\frac{1}{4}\sigma_{\mu\nu}\sigma^{\mu\nu} \, ,
\end{align*}

\begin{equation*}
	\mathcal{D}\equiv u^{\mu}\partial_{\mu},\quad 
		l^{\mu}=\epsilon^{\alpha\beta\gamma\mu}u_{\alpha}\partial_{\beta}u_{\gamma},
\end{equation*}

\begin{align*}
			\mathbf{v}_{4\,\nu} & =\frac{9}{5}\left[P^{\alpha}{}_{\nu}P^{\beta\gamma}\partial_{\gamma}\partial_{\left(\beta\right.}u_{\left.\alpha\right)}
				-\frac{1}{3}P^{\alpha\beta}P^{\gamma}{}_{\nu}\partial_{\gamma}\partial_{\alpha}u_{\beta}\right]
				-P^{\mu}{}_{\nu}P^{\alpha\beta}\partial_{\alpha}\partial_{\beta}u_{\mu},\\
			\mathbf{v}_{5\,\nu} & =P^{\mu}{}_{\nu}P^{\alpha\beta}\partial_{\alpha}\partial_{\beta}u_{\mu},\\
			\mathfrak{V}_{1\,\nu} & =\partial_{\alpha}u^{\alpha}\mathcal{D}u_{\nu},\quad
				\mathfrak{V}_{2\,\nu}=\epsilon_{\alpha\beta\gamma\mu}u^{\alpha}\mathcal{D}u^{\beta}l^{\gamma},\quad
				\mathfrak{V}_{3\,\nu}=\frac{1}{2}\sigma_{\alpha\nu}\mathcal{D}u^{\alpha}
\end{align*}
and $ \sigma_{\mu\nu} $, $ \Sigma_{\mu\nu}^{\left(i\right)} $ and $ \omega_{\mu\nu} $ were defined in (\ref{2orderTemrs}).

The functions $ \tilde{\mbox{s}}_{i}(\rho) $, $ \tilde{\mbox{k}}_{i}(\rho) $, $ \tilde{\mbox{w}}_{i}(\rho) $ and $ \tilde{\pi}_{i}(\rho) $ are determined by solving (\ref{E:SOL}). We have not written the explicit form of the sources $ \mathbf{S}^{(2)} $, $ \mathbf{k}^{(2)} $ and $ \mathbf{V}_{\mu}^{(2)} $. The sources for the second order contribution to $ \Pi_{\mu\nu} $, $ \mathbf{P}_{\mu\nu}^{(2)}, $ are given in Appendix \ref{appB}.

Once the functions $ \tilde{\mbox{s}}_{i}(\rho) $, $ \tilde{\mbox{k}}_{i}(\rho) $, $ \tilde{\mbox{w}}_{i}(\rho) $ and $ \tilde{\pi}_{i}(\rho) $ are determined, we can expand them around the asymptotic boundary,
\begin{align}
\label{E:Expansion}
\begin{split}
	S & =  S_{0}+\frac{b^{2}}{\rho^{2}}\left(\mbox{s}_{4}\,\mathfrak{S}_{4}
		+\mbox{s}_{5}\,\mathfrak{S}_{5}\right)\\
	k & =  -1+\frac{\kappa_{4}}{\rho^{4}}
		+b \frac{2S_{0}}{3\rho}\,\partial_{\alpha}u^{\alpha}
		+\frac{b^{2}}{\rho^{2}}\left(\mbox{k}_{0}\,\mathbf{s}_{3}
			+\mbox{k}_{1}\,\mathfrak{S}_{1}+\mbox{k}_{3}\,\mathfrak{S}_{3}
			+\mbox{k}_{4}\,\mathfrak{S}_{4}+\mbox{k}_{5}\,\mathfrak{S}_{5}\right)\\
	V_{\mu} & =  -b\frac{S_{0}}{\rho}\, u^{\alpha}\partial_{\alpha}u_{\mu}
		+\frac{b^{2}}{\rho^{2}}\left(\mbox{w}_{4}\,\mathbf{v}_{4\,\mu}
			+\mbox{w}_{5}\,\mathbf{v}_{5\,\mu}+\mbox{v}_{1}\,\mathfrak{V}_{1\,\mu}
			+\mbox{v}_{2}\,\mathfrak{V}_{2\,\mu}+\mbox{v}_{3}\,\mathfrak{V}_{3\,\mu}\right)\\
	\Pi_{\mu\nu} & =  b\left(\frac{\pi_{1}}{\rho}+\frac{\pi_{4}}{\rho^{4}}\right)\sigma_{\mu\nu}
		+b^{2}\sum_{i=0}^{3}\left\{ \left(\frac{\tilde{\pi}_{2\, i}}{\rho^{2}}
			+\frac{\tilde{\pi}_{4\, i}}{\rho^{4}}\right)\Sigma_{\mu\nu}^{\left(i\right)}\right\} \, ,
\end{split}
\end{align}
where
\begin{align} \label{E:element4}
	\kappa_{4}= \frac{S_{0}^{2}}{2S_{0}^{2}-1},\quad
		\pi_{1}=S_{0},\quad
		\pi_{4}=-\frac{S_{0}}{4} \left(2S_{0}^{2}-1\right)
\end{align}
are found using the zeroth (\ref{0th_order_functions}) and first (\ref{1st_order_functions}) order metric functions. The coefficients $ \pi_{i} $, $ \tilde{\pi}_{2\, i} $ and $ \tilde{\pi}_{4\, i} $ can be obtained by expanding (\ref{E:SOL}) near the boundary located at $ \rho \rightarrow \infty $. In particular, we find that
\begin{align} \label{E:Pi_tilde}
\begin{split}
	\tilde{\pi}_{4\, i} &= \frac{S_{0}^{2} \left( 2 S_{0}^{2} - 1 \right)}{24}
		\Bigg\{ \frac{6 \mathbf{P}_{i}^{(2)}(t)}{t} + 5 \mathbf{P}_{i}^{(2) \prime}(t) + t \, \mathbf{P}_{i}^{(2) \prime\prime}(t) \\
			& \qquad\qquad\qquad\qquad + \int_t^0 \left( \frac{6 \mathbf{P}_{i}^{(2) \prime}(z)}{z} + 6 \mathbf{P}_{i}^{(2) \prime\prime}(z)
					+ z \mathbf{P}_{i}^{(2) \prime\prime\prime}(z)\right) dz \Bigg\} \, ,
\end{split}
\end{align}
where each $ \mathbf{P}_{i}^{(2)} $ is a part of the following decomposition
\begin{equation} \label{2orderPIsources}
	\mathbf{P}^{(2)}_{\mu\nu} = \mathbf{P}_0^{(2)} \Sigma^{(0)}_{\mu\nu} + \mathbf{P}_1^{(2)} \Sigma^{(1)}_{\mu\nu} + \mathbf{P}_2^{(2)} \Sigma^{(2)}_{\mu\nu} + \mathbf{P}_3^{(2)} \Sigma^{(3)}_{\mu\nu} \, .
\end{equation}
The explicit form of the source terms $ \mathbf{P}_i^{(2)} $ ($ i=0,\ldots,3 $) appear in the Appendix \ref{appB}. Inserting (\ref{E:Expansion}) into (\ref{E:Tmnprescription}) we find
\begin{equation} \label{E:Tmn}
T^{\mu\nu} = P \left(4 u^{\mu}u^{\nu} + \eta^{\mu\nu}\right) - \eta \sigma^{\mu\nu} + \sum_{i=0}^{3} \lambda_{i} \Sigma^{(i)\mu\nu} \, ,
\end{equation}
where $ P $ and $ \eta $ were determined in (\ref{zeroth_order}) and (\ref{eta}) respectively and
\begin{equation} \label{E:TranspCoeffPrescription}
	\lambda_{i} = \frac{\tilde{\pi}_{4\, i}\left(2S_{0}^{2}-1\right)}{4\pi b^2 G_{5}S_{0}} \, .
\end{equation}

We have not managed to reduce (\ref{E:Pi_tilde}) to a closed form expression. We do note, however, that
\begin{equation}
	-2 \mathbf{P}_0^{(2)} + 4 \mathbf{P}_1^{(2)} - \mathbf{P}_2^{(2)} = \frac{d}{d\rho}\bm{\mathcal{S}}(\rho) \, ,
\end{equation}
where
\begin{align} \label{totalderiv}
\begin{split}
	\bm{\mathcal{S}}(\rho) =&
		-\frac{8\theta\left(2S_{0}^{4}+S_{0}^{2}\rho^{4}-4\theta\rho^{4}f\left(\rho\right)\right)}
			{S_{0}^{2}\left(1-8\theta\right)\rho^{2}\left(S_{0}^{2}-4\theta f\left(\rho\right)\right)}
		+\frac{4\left(S_{0}^{2}-S_{0}^{2}\rho^{3}-4\theta f\left(\rho\right)\right)}
			{S_{0}\left(S_{0}^{2}-4\theta f\left(\rho\right)\right)}\pi\left(\rho\right) \\
		&-\frac{8\theta\rho\left(2S_{0}^{2}-f\left(\rho\right)\right)\left(S_{0}^{2}\left(1+\rho^{3}\right)
			-4\theta f\left(\rho\right)\right)}{S_{0}^{3}\left(1-8\theta\right)\left(S_{0}^{2}-4\theta f\left(\rho\right)\right)}\pi'\left(\rho\right) \, .
\end{split}
\end{align}
Plugging (\ref{totalderiv}) into (\ref{E:SOL}) and then using (\ref{E:TranspCoeffPrescription}), we find that
\begin{equation} \nonumber
	- 2 \lambda_0 + 4 \lambda_1  - \lambda_2 = \frac{\pi^{2}T^{2}}{\pi G_{5}}\left[\frac{\sqrt{2}\left(1-\sqrt{1-8\theta}+16\theta\right)\theta}{\left(1+\sqrt{1-8\theta}\right)^{3/2}}\right] \,,
\end{equation}
as advertised in \eqref{FinResult}. This is the main result of our paper.

We note in passing that the authors of \cite{Andrei} have found closed form expressions for the $ \lambda_i $'s. We have compared our integral expression (\ref{E:Pi_tilde}) with their results both numerically and analytically and found excellent agreement. For instance, expanding (\ref{E:Pi_tilde}) at small $ \theta $ we get
\begin{align} \nonumber
\begin{split}
	\lambda_{0} & = \frac{\pi^2 T^2}{32 \pi G_5} \left[2-\log 2 + \left(-21+5 \log 2\right) \theta + 
		\left(5+\frac{21}{2}\log 2\right) \theta^2 + \mathcal{O}(\theta^3) \right]\\
	\lambda_{1} & = \frac{\pi^2 T^2}{32 \pi G_5} \left[1 - 7 \theta + \frac{151}{2} \theta^2 + \mathcal{O}(\theta^3) \right]\\
	\lambda_{2} & = \frac{\pi^2 T^2}{32 \pi G_5} \left[2 \log 2 + \left(14-10 \log 2\right) \theta -
		\left(28+21\log 2\right) \theta^2 + \mathcal{O}(\theta^3) \right]\\
	\lambda_{3} & = \frac{\pi^2 T^2}{32 \pi G_5} \left[- 112 \theta - 80 \theta^2 + \mathcal{O}(\theta^3) \right] \, ,
\end{split}
\end{align}
with
\begin{equation} \nonumber
	- 2 \lambda_0 + 4 \lambda_1  - \lambda_2 = \frac{\pi^2 T^2}{32 \pi G_5} 320 \, \theta^2 + \mathcal{O}(\theta^3) \, .
\end{equation}

\section*{Acknowledgements}
I would like to thank A. Yarom for useful discussions and collaboration. This work was supported by the ISF under grant numbers 495/11,
630/14 and 1981/14, by the BSF under grant number 2014350, by the European commission
FP7, under IRG 908049 and by the GIF under grant number 1156/2011.

\begin{appendices}

\section{Hawking black hole temperature}
To find the black-hole temperature we convert our ansatz (\ref{eq:ansatz0}) into a non-zero temperature black 3-brane metric. With $ dt \rightarrow dt - \frac{S_0}{r^2 f(br)}dr $, one gets
\begin{equation}
	ds^2 = -r^2 f(br) dt^2 + \frac{dr^2}{r^2 f(br)/S_0^2} + r^2 \left(dx^i\right)^2 \,.
\end{equation}
Since the Hawking temperature is completely determined at the vicinity of horizon, we expand the metric functions near $ r=r_h $ (in our case $ r_h = b^{-1} $ with $ f(b r_h) = 0 $)
\begin{equation} \nonumber
	ds^2 = -\left. (r^2 f(br))'\right|_{r=r_h} (r-r_h) dt^2 + 
		\frac{S_0^2 \, dr^2}{\left. (r^2 f(br))'\right|_{r=r_h} (r-r_h)} + r^2 \left(dx^i\right)^2 \, .
\end{equation}
Defining $ r-r_h = z^2 \frac{\left. (r^2 f(br))'\right|_{r=r_h}}{4 S_0^2} $ and using the zeroth order solution (\ref{0th_order_functions}), we find
\begin{align} \nonumber
\begin{split}
	ds^2 & =  - \frac{4 a^{1/2} S_0^2}{b^2} z^2 dt^2 + dz^2 + r^2 \left(dx^i\right)^2 \, .
\end{split}
\end{align}
After a Wick rotation to Euclidean time $ \tau $, we have (without the last term) a flat space metric in cylindrical coordinates $ ds^2 = dz^2 + z^2 d\varphi^2 $ where $ \varphi = \frac{2 S_0}{b} \tau $. To avoid a conical singularity at $ z=0 $, $ \varphi $ must have a periodicity of $ 2\pi $. Since periodicity of Euclidean time is the inverse temperature, we have $ 2\pi = \frac{2 S_0}{bT} $ and therefore
\begin{equation} \label{TEMP}
T = \frac{S_0}{b \pi} \, .
\end{equation}

\section{Second order source terms}\label{appB}
Source terms for the second order $ \Pi_{\mu\nu}^{(2)} $ metric functions, see (\ref{2orderPIsources}) and (\ref{E:SOL}).

\begin{align} \nonumber
\begin{split}
	\mathbf{P}_0^{(2)} & = \frac{S_0^6 \left(24 \, \theta + (1-8 \theta) \rho^4\right)}{\rho^3 \left(S_0^2-4\theta f(\rho)\right)^3}
		+ \frac{S_0^5 \left(-40 \, \theta - 3 (1-8\theta)\rho^4\right)}{\rho^2 \left(S_0^2-4\theta f(\rho)\right)^3} \, \pi(\rho) + \frac{-2 S_0 \rho^3}{S_0^2-4\theta f(\rho)} \pi'(\rho)
\end{split}
\end{align}
\begin{align*} \nonumber
	\mathbf{P}_1^{(2)}  = &\frac{\rho}{S_0^{2}}\left(\frac{-4\theta}{1-8\theta}
		+\frac{-2 S_0^{8}\left(1-8\theta\right)}{\left(S_0^{2}-4\theta f(\rho)\right)^{3}}
		+\frac{3 S_0^{4}}{S_0^{2}-4\theta f(\rho)}\right) +\frac{-S_0^{5}\left(40\theta+3\left(1-8\theta\right)\rho^{4}\right)}
			{\rho^{2}\left(S_0^{2}-4\theta f(\rho)\right)^{3}} \pi(\rho) \\ 
		&+\frac{2\rho^{3}}{S_0}\left(\frac{-2}{1-8\theta}
		+\frac{-S_0^{6}\left(1-8\theta\right)}{\left(S_0^{2}-4\theta f(\rho)\right)^{3}}
		+\frac{3S_0^{2}}{S_0^{2}-4\theta f(\rho)}\right)\pi'(\rho) \\
		&+\frac{\rho^{5}\left(-16S_0^{8}\theta+f(\rho)\left(S_0^{6}-4\theta f(\rho)\left(S_0^{2}\left(1-16\theta\right)
		+4\theta f(\rho)\right)\left(3S_0^{2}-8\theta f(\rho)\right)\right)\right)}{S_0^{2}\left(1-8\theta\right)\left(S_0^{2}
		-4\theta f(\rho)\right)^{3}}\left(\pi'(\rho)\right)^{2}\\
		&+\frac{\rho^{4}}{S_0}\left(\frac{-1}{1-8\theta}+\frac{S_0^{2}}{S_0^{2}-4\theta f(\rho)}\right)\pi''(\rho)
		+\frac{-4\theta\rho^{6}f(\rho)\left(2S_0^{2}-f(\rho)\right)}{S_0^{2}\left(1-8\theta\right)\left(S_0^{2}
		-4\theta f(\rho)\right)}\pi'(\rho)\pi''(\rho)
\end{align*}
\begin{align} \nonumber
\begin{split}
	\mathbf{P}_2^{(2)} & = \frac{2 S_0^2 \rho}{S_0^2-4\theta f(\rho)}
		+ \frac{2 S_0^5 \left(40 \, \theta + 3 (1-8\theta)\rho^4\right)}{\rho^2 \left(S_0^2-4\theta f(\rho)\right)^3} \, \pi(\rho) + \frac{4 S_0 \rho^3}{S_0^2-4\theta f(\rho)} \pi'(\rho)
\end{split}
\end{align}
\begin{align} \nonumber
\begin{split}
	\mathbf{P}_3^{(2)} & = \frac{S_0^2 \left(24 \, \theta +(1-8 \theta) \rho^4\right)}{\theta (1-8 \theta) \rho^3}
		+ \frac{S_0^6 \, (1-8\theta)\, (3 S_0^2 - 16\theta) \rho}{\theta \left(S_0^2-4\theta f(\rho)\right)^3} 
		+ \frac{4 S_0^2 \, (6\theta - S_0^2) \rho}{\theta \left(S_0^2-4\theta f(\rho)\right)}
\end{split}
\end{align}

\end{appendices}

\addcontentsline{toc}{section}{Bibliography}
\bibliographystyle{JHEP}   
\bibliography{mybibfile}      


\end{document}